# Impact of ADAS and V2X Penetration Rates on Cooperative Active Safety

M.C. Lucas-Estañ[1], B. Coll-Perales[1], M. I. Khan[2], J. Gozalvez[1], O. Altintas[2]
[1]Uwicore laboratory, Universidad Miguel Hernandez de Elche, Elche (Alicante), Spain.
[2]InfoTech Labs, Toyota Motor North America R&D, Mountain View, CA, U.S.A.
{m.lucas, bcoll, j.gozalvez}@umh.es; {mohammad.irfan.khan, onur.altintas}@toyota.com

***Abstract*— A major driver of connected and automated driving is cooperative active safety. The effectiveness of cooperative safety applications depends on the ability of vehicles to detect traffic safety risks in advance. Such risks can be identified either through ADAS (Advanced Driving Assistance Systems) or via V2X (Vehicle to Everything) communications. More vehicles are gradually being deployed with ADAS, but ADAS sensors can be limited by their sensing range and field of view. On the other hand, V2X can experience communication ranges beyond the ADAS sensing range, but its impact is highly dependent on the V2X penetration rate. This paper analyzes the impact of ADAS and V2X penetration rates on the effectiveness of cooperative active safety applications considering an emergency braking maneuver use case in a highway scenario. Results show that while ADAS and V2X each enhance traffic safety, their combined deployment further amplifies these gains, with the effect becoming more pronounced as V2X is deployed more rapidly.**

## I. Introduction

The advancement of connected and automated driving is largely driven by the pursuit of active traffic safety. Cooperative safety applications aim to anticipate and mitigate accident risks before they lead to collisions. ADAS (Advanced Driving Assistance Systems) and V2X (Vehicle to Everything) communications can serve as fundamental enablers, extending the driver's perception and decision-making beyond human limitations and providing early alerts when critical driving situations arise. Vehicles are increasingly equipped with ADAS systems that rely on on-board sensors such as cameras, LiDAR, and radars for automatic detection of surrounding objects and potential safety hazards. V2X can complement ADAS as it increases the vehicles' awareness range and can provide notifications from vehicles beyond the sensor field of view. However, its benefits are strongly conditioned by the penetration rate as all vehicles involved in the safety risk must embed V2X technologies. In contrast, ADAS can provide safety benefits in certain scenarios even if only the ego vehicle is equipped with ADAS sensors, e.g. in an emergency braking maneuver. These benefits would be augmented if vehicles were equipped with V2X as an earlier notification of the emergency braking maneuver could be obtained leading to a smoother deceleration and safer maneuver, in particular when several vehicles are involved in the maneuver.

Several studies have already shown that vehicles equipped with ADAS (e.g., [1]) or V2X (e.g., [2]) can improve safety, with most studies focusing on intersections due to the challenges caused by building obstructions between approaching vehicles. Studies usually consider scenarios in which both vehicles approaching the intersection are equipped with V2X or ADAS and analyze whether timely detection is feasible under different setups, such as vehicle speed, ADAS sensing capabilities, or V2X configuration (e.g., transmission power and rate). The impact of V2X penetration rates in intersection scenarios is analyzed in [3], where the authors show that collisions decrease by more than 20% only when all vehicles are equipped with V2X. Recent studies [4]-[6] evaluate the safety gains achieved with both ADAS and V2X in emergency braking at intersections. The authors design a two-stage braking strategy consisting of a partial brake triggered by V2X followed by ADAS-based emergency braking. The studies assume vehicles are automated, and all vehicles involved in the maneuver embed both technologies. However, safety gains can also be achieved even if vehicles are not (fully) automated, and the gains may strongly depend on the ADAS and V2X penetration rates.

This study extends the state-of-the-art by analyzing the impact of ADAS and V2X penetration rates in cooperative active safety considering an emergency braking use case, as it represents one of the main causes of critical traffic situations according to recent reports [7]. The evaluation considers scenarios where vehicles are equipped with one of the two technologies, both, or neither, without assuming full automation. The study introduces a methodology to assess not only the impact on crash avoidance, but also on accident severity and the conditions under which collisions can be prevented. Furthermore, the methodology and analysis account for chain effects in emergency braking maneuvers, an aspect typically overlooked despite its significant influence on safety in such scenarios. The analysis shows that both ADAS and V2X reduce the probability of collisions during emergency braking maneuvers with the gains increasing with the penetration rate. V2X penetration rates above 20% are necessary in this use case to achieve greater safety benefits than with ADAS alone. However, safety benefits are significantly amplified when vehicles are equipped with both V2X and ADAS, advocating for a faster deployment of V2X.

## II. Emergency Braking Analysis

We consider an emergency braking use case where a front vehicle suddenly decelerates and may cause rear-end

This work was supported in part by MICIU/AEI/10.13039/501100011033 (grant ID2023-150308OB-I00), by Generalitat Valenciana, and UMH's Vicerrectorado de Investigación.

collisions. We distinguish between the case where a rear-end collision is caused by the vehicle directly in front of the ego vehicle (rear-end collision illustrated in Fig. 1.a), and the case of potential chain collisions when the sudden deceleration is caused by other front vehicles (chain collisions illustrated in Fig. 1.b and Fig. 1.c). A rear-end collision can occur if the following vehicle detects the deceleration too late and/or does not have sufficient time to react and avoid the collision. Chain collisions occur when the sudden braking maneuver from the front vehicle triggers a cascading sequence of delayed reactions along following vehicles. Each driver reacts with some delay relative to the preceding vehicle, and these delays accumulate downstream. While the first vehicle following the front vehicle may avoid a collision, the growing cumulative response time significantly increases the likelihood of multi-vehicle rear-end collisions, which are common in highway traffic.

This study assumes all vehicles are controlled by a driver who performs the braking maneuver to avoid a rear-end collision. In this case, the ability of an ego vehicle to avoid a rear-end collision depends on what we define as the response time ($\rho$), which is the time elapsed between the instant when the leading vehicle performs an emergency braking maneuver and the moment when the driver at the ego vehicle perceives (or is informed of) the maneuver and reacts.

### A. Response time

The response time is equal to the driver's reaction time for vehicles not equipped with ADAS or V2X ($\rho_U$), i.e., unequipped vehicles. The driver's reaction time [8] is the sum of the perception time and the motor response time. Perception time refers to the time required for the driver to become aware of a traffic situation (e.g., a vehicle braking ahead). Motor response time is the time needed for the driver to initiate a maneuver after becoming aware of the traffic situation (e.g., pressing the brake pedal) and for the vehicle's actuators to produce the corresponding response (e.g., decelerating the vehicle). The motor response time is equal for all vehicles independently of whether they are equipped with ADAS and/or V2X or not, since we assume all vehicles are controlled by the driver. On the other hand, the perception time is different for vehicles without ADAS and V2X and vehicles with ADAS and/or V2X. This results in different reaction times for unequipped vehicles and vehicles equipped with ADAS and/or V2X.

The response time when the vehicle that brakes is detected by ADAS ($\rho ADAS$) is the sum of the driver's reaction time and the time required for the on-board sensors to detect the object that triggers the event notification to the driver through the HMI ($Ldet$). $Ldet$ depends on the detection algorithm and the rate at which detection outputs are generally delivered by ADAS, which typically ranges from 10 Hz to 50 Hz [9][10]. The reaction time is again equal to the perception time and the motor response time. However, vehicles equipped with ADAS experience a lower perception time than unequipped vehicles as notifications are delivered to the HMI as soon as a risk is detected. The use of simple stimuli such as visual alerts has been shown to reduce a driver's perception time [11].

The response time when the vehicle that brakes is detected by V2X is equal to the sum of the reaction time, the time the vehicle ahead that brakes needs to generate the V2X message informing about the deceleration ($Lgen$), and the V2X communication latency ($Lcom$). Vehicles equipped with V2X also use an HMI to inform the driver about an event notification, and hence reduce the reaction time with respect to unequipped vehicles.

Vehicles equipped with ADAS and V2X can respond to events detected by either technology. Consequently, their response time is the minimum of $\rho_{ADAS}$ and $\rho_{V2X}$.

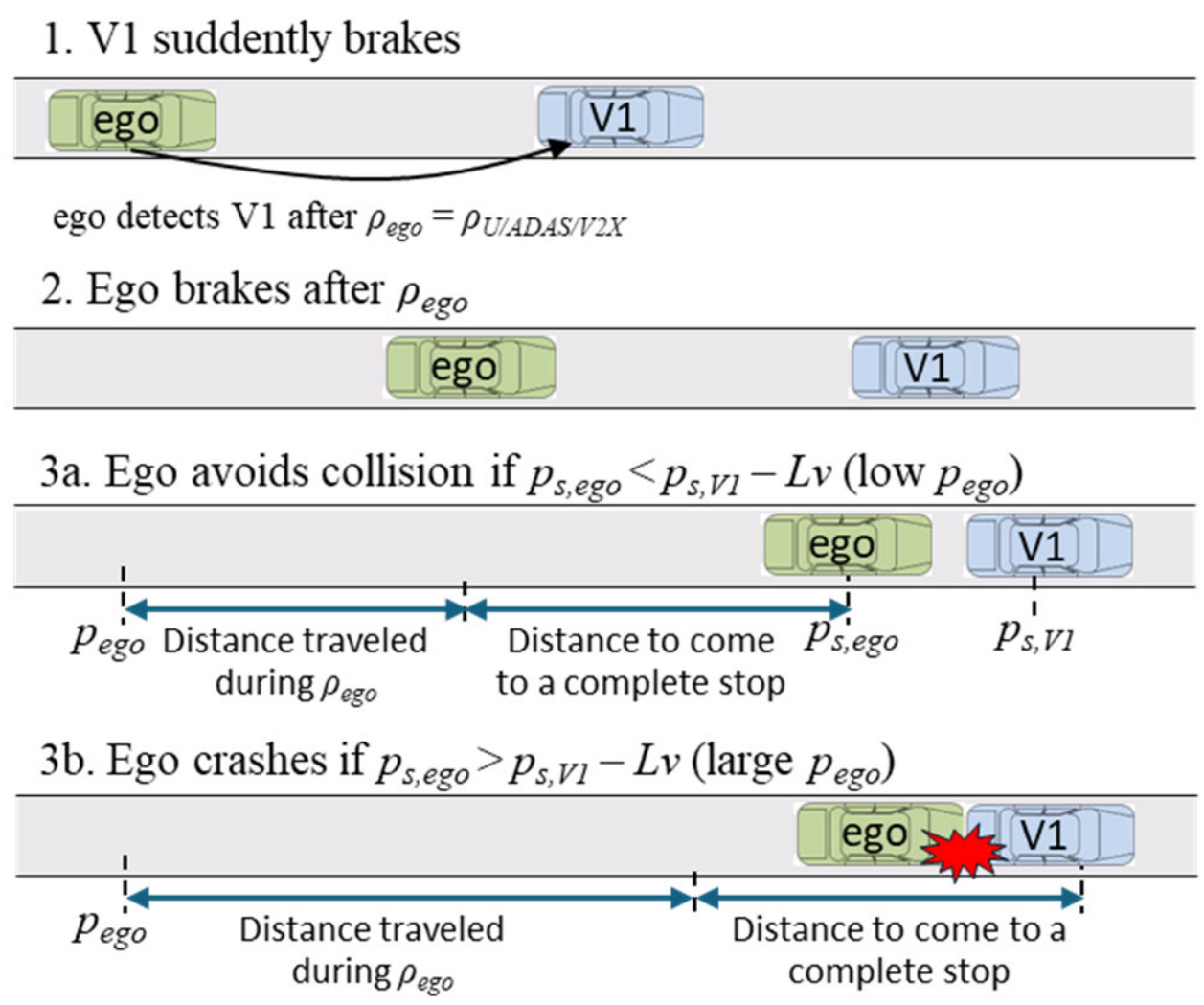

a) Rear-end collision.

1a. V2 suddenly brakes. Ego detects V2 via V2X

V1 detects V2 after $\rho_{V1} = \rho_{U/ADAS/V2X}$

$p_{ego}$ $p_{V1}$ $p_{V1}$

ego detects V2 after $\rho_{ego} = \rho_{V2X}$

1b. V2 suddenly brakes. Ego does not detect V2 but detects V1

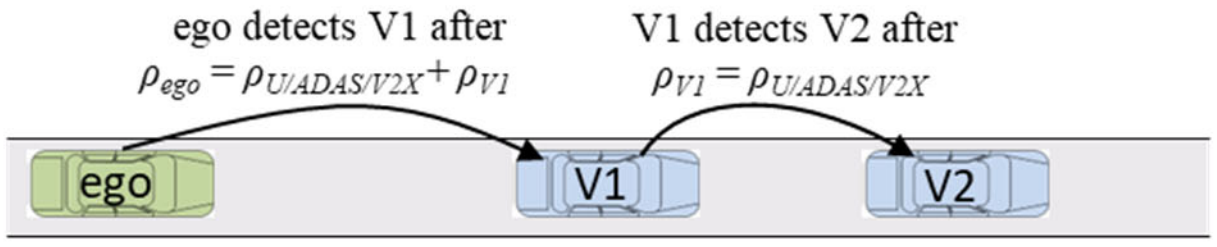

2. V1 comes to a complete stop just before collison with V2

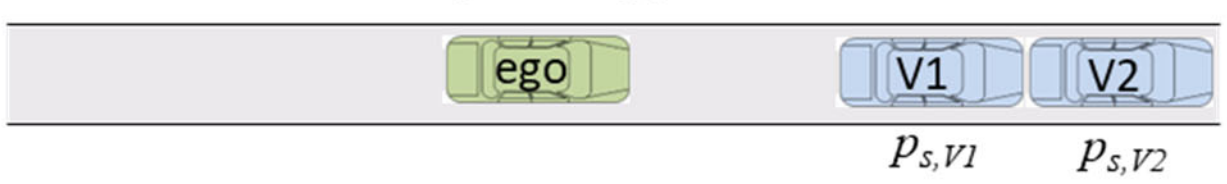

3a. Ego avoids collision if $p_{s,ego} < p_{s,V1} - Lv$ (low $\rho_{ego}$)

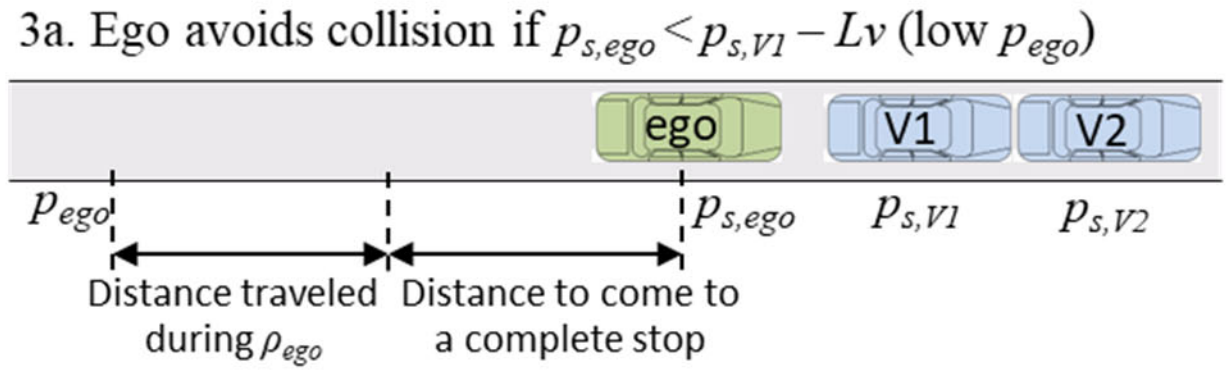

3b. Ego crashes if $p_{s,ego} > p_{s,V1} - Lv$ (large $\rho_{ego}$)

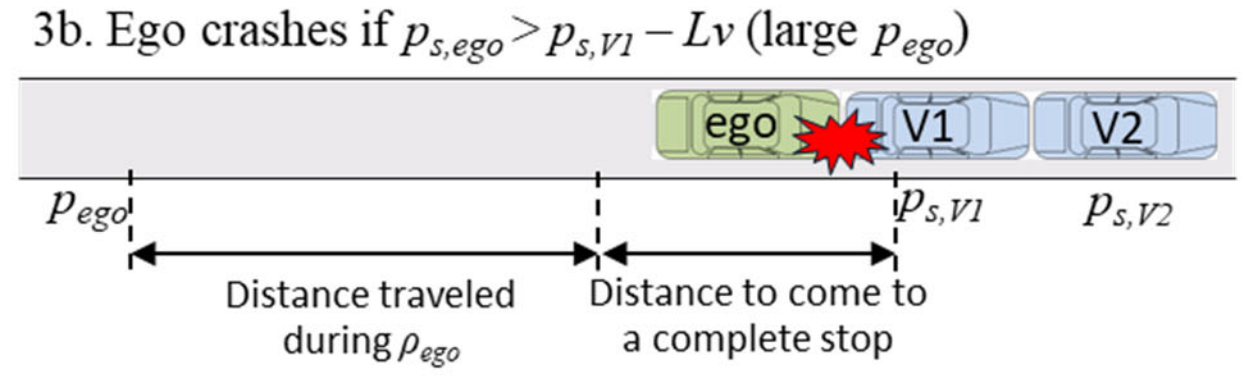

c) Chain collision.

Fig. 1. Examples of rear-end and chain collisions cases.

## B. Rear-end collision

Fig. 1.a represents the case of a *direct rear-end collision* when $V_1$, that is at position $p_{V1}$ moving at speed $v_{V1}$, suddenly brakes at $t$=0 with acceleration $a_{V1}$ = -9 m/s$^2$ and comes to a complete stop at position $p_{s,V1}$. We assume that the ego vehicle that is at position $p_{ego}$ moves at a constant speed $v_{ego}$ during the driver's response time $\rho_{ego}$. The value of $\rho_{ego}$ depends on whether vehicles involved in the safety risk embed ADAS, V2X, both technologies or none. In particular, $\rho_{ego}$ is determined as follows:

- $\rho_{ego} = \rho_U$ if the ego vehicle does not have ADAS or V2X.
- $\rho_{ego} = \rho_{ADAS}$ if the ego vehicle has ADAS and $V_1$ is within the ADAS sensing range, otherwise, $\rho_{ego} = \rho_U$.
- $\rho_{ego} = \rho_{V2X}$ if both the ego vehicle and $V_1$ have V2X, and the ego vehicle successfully receives a notification from $V_1$ when it suddenly brakes. If $V_1$ does not have V2X or the notification message is lost due to a V2X transmission error, $\rho_{ego} = \rho_U$.
- If the ego vehicle has both ADAS and V2X, $\rho_{ego} = \rho_{ADAS}$ if the sudden brake is detected by ADAS and not received through V2X because $V_1$ does not have V2X or the notification was not correctly received. If the V2X notification is received and the deceleration was not detected by ADAS, then $\rho_{ego} = \rho_{V2X}$. If the deceleration is detected with ADAS and V2X, then $\rho_{ego}$ = min($\rho_{ADAS}$, $\rho_{V2X}$). Otherwise, $\rho_{ego} = \rho_U$.

We assume the ego vehicle begins the emergency braking at time $\rho_{ego}$ with acceleration $a_{ego}$ = -9 m/s$^2$, and would come to a complete stop at position $p_{s,ego}$ if no collision with $V_1$ occurs. The ego vehicle collides with $V_1$ if $p_{s,ego} > p_{s,V1}-L_V$, where $L_V$ is the vehicle length. In case of collision, Algorithm I determines the positions and speeds of $V_1$ and the ego vehicle ($p_{col,ego}$, $p_{col,V1}$, $v_{col,ego}$, $v_{col,V1}$) at the time of the collision. These parameters are relevant to quantify the severity of collisions.

## C. Chain collisions

We now explain how the evaluation of emergency braking maneuvers can be extended to a *chain collision* scenario. We consider a simple scenario where the ego vehicle might collide with $V_1$ when $V_2$ performs a sudden deceleration, but the methodology can be extended to any number of vehicles involved in *chain collisions*; this has been the case for our numerical evaluation. *Chain collisions* are evaluated considering that $V_2$ suddenly brakes at $t$ = 0 with $a_{V2}$ = -9 m/s$^2$, and comes to a complete stop at position $p_{s,V2}$. We assume that $V_1$ moves at constant speed $v_{V1}$ during the driver's response time $\rho_{V1}$. After $\rho_{V1}$, $V_1$ brakes with the minimum deceleration required to avoid a collision with $V_2$, limited to -9 m/s$^2$. Thus, $a_{V1}$ = max($a_{V1,min}$, -9) m/s$^2$, where $a_{V1,min}$ is the acceleration necessary for $V_1$ to stop just before colliding with $V_2$. If $a_{V1,min}$ < -9 m/s$^2$, $a_{V1}$ is set equal to -9 m/s$^2$, and $V_1$ will collide with $V_2$. The position of $V_1$ at the moment of impact ($p_{col,V1}$) can be determined using Algorithm I. If $V_1$ will not collide with $V_2$, $V_1$ comes to a complete stop at $p_{s,V2}$. The ego vehicle continues at constant speed $v_{ego}$ until it detects and reacts to the emergency braking situation after the response time $\rho_{ego}$. $\rho_{ego}$ is determined as follows:

- If the ego vehicle does not have ADAS or V2X, it can only react after detecting that $V_1$ is braking since $V_2$ is visually obstructed by $V_1$. In this case, $\rho_{ego} = \rho_U + \rho_{V1}$.

**Algorithm I: Collision severity metrics**

Input: $p_{ego}$, $v_{ego}$, $a_{ego}$, $\rho_{ego}$, $p_{V1}$, $v_{V1}$, $a_{V1}$
Output: $v_{col,ego}$, $v_{col,V1}$, $p_{col,ego}$, $p_{col,V1}$
1. If collision occurs before V1 starts braking
2. Calculate $t_{col}$ that satisfies: $(a_{V1}/2)\cdot t_{col}{}^2 + (v_{V1}-v_{ego})\cdot t_{col} + p_{V1} - p_{ego} - L_v = 0$
3. $v_{col,V1} = \max(v_{V1} + a_{V1}\cdot t_{col}, 0)$, $v_{col,ego} = v_{ego}$
4. $p_{col,ego} = p_{ego} + v_{ego}\cdot t_{col}$, $p_{col,V1} = p_{col,ego} + L_v$
5. Else
6. If collision occurs after V1 comes to a complete stop
7. $p_{col,V1} = p_{V1} - 3/2\cdot v_{V1}{}^2/a_{V1}$, $p_{col,ego} = p_{col,V1} - L_v$
8. Calculate $t_{col}$ that satisfies: $(a_{ego}/2)\cdot t_{col}{}^2 + (v_{ego} - a_{ego}\cdot\rho_{ego})\cdot t_{col} + p_{ego} - v_{ego}\cdot\rho_{ego} + a_{ego}/2\cdot\rho_{ego}{}^2 - p_{ego.col} = 0$
9. $v_{col,ego} = v_{ego} + a_{ego}\cdot t_{col}$, $v_{col,V1} = 0$
10. Else
11. Calculate $t_{col}$ that satisfies: $(a_{V1} - a_{ego})/2\cdot t_{col}{}^2 + (v_{V1} - v_{ego} + a_{ego}\cdot t_{col}\cdot\rho_{ego})\cdot t_{col} + p_{V1} - p_{ego} - a_{ego}/2\cdot\rho_{ego}{}^2 - L_v = 0$
12. $v_{col,V1} = v_{V1} + a_{V1}\cdot t_{col}$, $v_{col,ego} = v_{ego} + a_{V1}\cdot(t_{col} - \rho_{ego})$
13. $p_{col,ego} = p_{V1} + v_{V1}\cdot t_{col} + a_{V1}/2\cdot t_{col}{}^2 - L_v$, $p_{col,V1} = p_{col,ego} + L_v$
14. End
15. End

- Even if the ego vehicle has ADAS, it can only react after detecting that $V_1$ is braking since $V_1$ obstructs the ADAS detection of $V_2$. In this case, $\rho_{ego} = \rho_{ADAS} + \rho_{V1}$ if $V_1$ is detected via ADAS. Otherwise, $\rho_{ego} = \rho_U + \rho_{V1}$.
- If the ego vehicle is equipped with V2X, it can receive from $V_2$ a notification of its sudden deceleration if $V_2$ is also equipped with V2X and the event-triggered V2X message is correctly received. In this case, $\rho_{ego} = \rho_{V2X}$. Otherwise, if the ego vehicle receives a V2X message from $V_1$ notifying that $V_1$ is braking, $\rho_{ego} = \rho_{V2X} + \rho_{V1}$. If neither message is received, $\rho_{ego} = \rho_U + \rho_{V1}$.
- If the ego vehicle is equipped with both ADAS and V2X, $\rho_{ego}$ can be calculated as $\rho_{ego} = \min(\rho_{ego\text{-}ADASonly}, \rho_{ego\text{-}V2Xonly})$, where $\rho_{ego\text{-}ADASonly}$ and $\rho_{ego\text{-}V2Xonly}$ are the response time assuming that the ego vehicle is equipped with ADAS only and V2X only, respectively.

$V_{ego}$ would come to a complete stop at position $p_{s,ego}$ if no collision with $V_1$ occurs. The ego vehicle collides with $V_1$ if $p_{s,ego} > p_{col,V1}-L_V$ in case $V_1$ collided with $V_2$, or if $p_{s,ego} > p_{s,V1}-L_V$ otherwise. The positions and speeds of $V_1$ and the ego vehicle at the time of the collision can be calculated extending Algorithm I to account for additional situations that may occur in chain collisions, for example, the possibility that a collision between the ego vehicle and $V_1$ occurs before or while $V_1$ is braking as a result of the deceleration of $V_2$.

# III. EVALUATION SCENARIO

We utilize SUMO to generate realistic mobility traces and assess potential safety risks if vehicles in the scenario suddenly brake. We consider a 5 km highway segment with two lanes in each direction. Vehicles travel at an average speed of 130 km/h, and the traffic density is 60 veh/km. We analyze the occurrence of rear-end collisions when a vehicle in the scenario suddenly brakes with emergency acceleration of -9 m/s$^2$.

We analyze scenarios where vehicles are equipped with ADAS only, V2X only or with both technologies, and analyze performance under different penetration rates of the technologies. The driver's reaction for unequipped vehicles is set equal to 2.5 s and equal to 0.75 s for vehicles equipped with ADAS and/or V2X due to their lower perception time [11]. The time *Ldet* between ADAS detection updates is set to 100 ms [9][10]. The ADAS field of view is modeled as a sensor cone with a 120 m sensing range and 120° angular aperture as in [4]. We implement a function to identify potential blockage from surrounding vehicles, and the ADAS can detect other vehicles that fall within its sensing range as long as they are not obstructed by other vehicles between them.

Without loss of generality, we assume that an emergency braking maneuver automatically generates an event-notification V2X message, as in the case of (Decentralized Environmental Notification Message) DENM [12], and set *Lgen* equal to 10 ms. The V2X communication latency *Lcom* is derived using an empirical latency cumulative distribution function (CDF) obtained with an ns-3–based 5G-NR V2X Mode 2 simulator [13]. The communications latency is obtained setting NR V2X mode 2 with $T_1$=2.5 ms and *PDB*=100 ms. $T_1$ is the processing time required to identify candidate resources within the selection window for transmitting a V2X packet, and *PDB* represents the Packet Delay Budget, i.e., the latency deadline by which the V2X packet must be transmitted. V2X transmission errors are modeled using an analytical 5G NR V2X mode 2 model derived from the C-V2X model presented in [14]. The model quantifies the packet delivery ratio (PDR) in V2X communications as a function of the distance between communicating vehicles. The model accounts for path loss and shadowing effects under line-of-sight (LOS) and non-line-of-sight conditions due to vehicle blockage (NLOSv) in accordance with 3GPP TR 37.885. It accounts for multi-path fading effects using link-level curves that model the Block Error Rate (BLER) as a function of the Signal-to-Interference-plus-Noise Ratio (SINR) for different 5G NR V2X Modulation and Coding Schemes (MCSs) [15]. It also accounts for half-duplex constraints and possible packet collisions in a 20 MHz channel. In addition to DENMs, vehicles transmit periodic 400 bytes packets at 10 Hz (e.g. Cooperative Awareness Message -CAM-), using MCS13, a transmission power of 23 dBm, and numerology 1.

The configuration parameters are summarized in Table I.

TABLE I. CONFIGURATION PARAMETERS

| **Parameter** | **Value** |
|---|---|
| Average speed | 130 km/h |
| Traffic density | 60 veh/km |
| Emergency acceleration | -9 m/s$^2$ |
| Reaction time | 2.5 s (unequipped vehicle), 0.75 s (equipped vehicle) |
| *Ldet* | 100 ms |
| *Lgen* | 10 ms |
| *Lcom* | $U(T_1$, PDB) |
| *PDB* | 100 ms |
| Sensing range and angle | 120 m and 100º |
| 5G (PC5) bandwidth | 20 MHz |
| Transmission power | 23 dBm |
| 5G numerology | 1 |
| Packet size | 400 bytes |
| Packet rate | 10 Hz |
| MCS | 13 |

## IV. RESULTS

Fig. 2 depicts the percentage of collisions avoided when deploying ADAS and V2X technologies compared to the scenario where none of these technologies are deployed. The performance is shown for scenarios where vehicles are equipped only with ADAS (*ADAS*), only with V2X (*V2X*), and with both technologies (*ADAS+V2X*). The performance is shown as a function of the penetration rate of the technologies over different years (*Y*) and includes rear-end and chain collisions. Fig. 2.a corresponds to a scenario where ADAS and V2X are deployed at equal penetration rates, starting from 5% to 100%. Fig. 2.b and Fig. 2.c correspond to more probable scenarios with higher ADAS penetration rates than V2X. Fig. 2.a shows that both ADAS and V2X improve safety and reduce the number of potential collisions during emergency braking maneuvers, with benefits increasing with the penetration of the technologies. ADAS provides greater safety gains than V2X at low penetration rates because the detection of the emergency braking depends only on the sensing capabilities of the ego vehicle. In contrast, the detection with V2X requires that both the braking vehicle and the ego vehicle be equipped with V2X so that they can exchange the notification. Consequently, the probability of detecting a braking vehicle using V2X increases approximately with the square of the V2X penetration rate, whereas for ADAS it scales roughly linearly with its penetration rate. As penetration increases, the safety benefits of V2X augment and outperform ADAS because vehicles equipped with V2X can detect an emergency braking event and react earlier than vehicles relying solely on ADAS, particularly in the scenario of chain collisions. In such situations, vehicles with only ADAS can solely detect the immediately preceding vehicle, and the ego vehicle begins braking only after the sum of the intervening vehicles' response times and its own reaction time (Section II). In contrast, vehicles with V2X can also detect braking vehicles that are two or more positions ahead, provided those vehicles are also equipped with V2X, which is why the benefit of V2X over ADAS augments with medium and large penetration rates.

Fig. 2.a shows that the joint deployment of V2X and ADAS amplifies the safety benefits and significantly reduces the number of collisions. For instance, the number of collisions is reduced by half when 45% of vehicles are equipped with both technologies, while such gains with only V2X or ADAS would require penetration rates of about 55% and 66%, respectively. Fig. 2.a further shows that equipping vehicles with both technologies provides safety benefits even at very low penetration rates since it is possible to leverage the strengths of each technology.

Fig. 2.a demonstrates how the combined deployment of ADAS and V2X can amplify safety benefits. However, assuming equal penetration rates for both technologies is not realistic: current reports (e.g., [7]) show that ADAS is already present in over 30% of the global vehicle fleet, whereas V2X deployment is only beginning. We then evaluate the performance under scenarios with higher ADAS penetration rates and a *conservative* (Fig. 2.b) or *expanded* (Fig. 2.c) V2X deployment from year 10. For Y1–Y9, the ADAS penetration values are forecasted for future years by extrapolating the historical penetration rate of ADAS equipped vehicles [7]. Fig. 2.b and Fig. 2.c show that, despite the substantial difference between ADAS and V2X penetration levels,

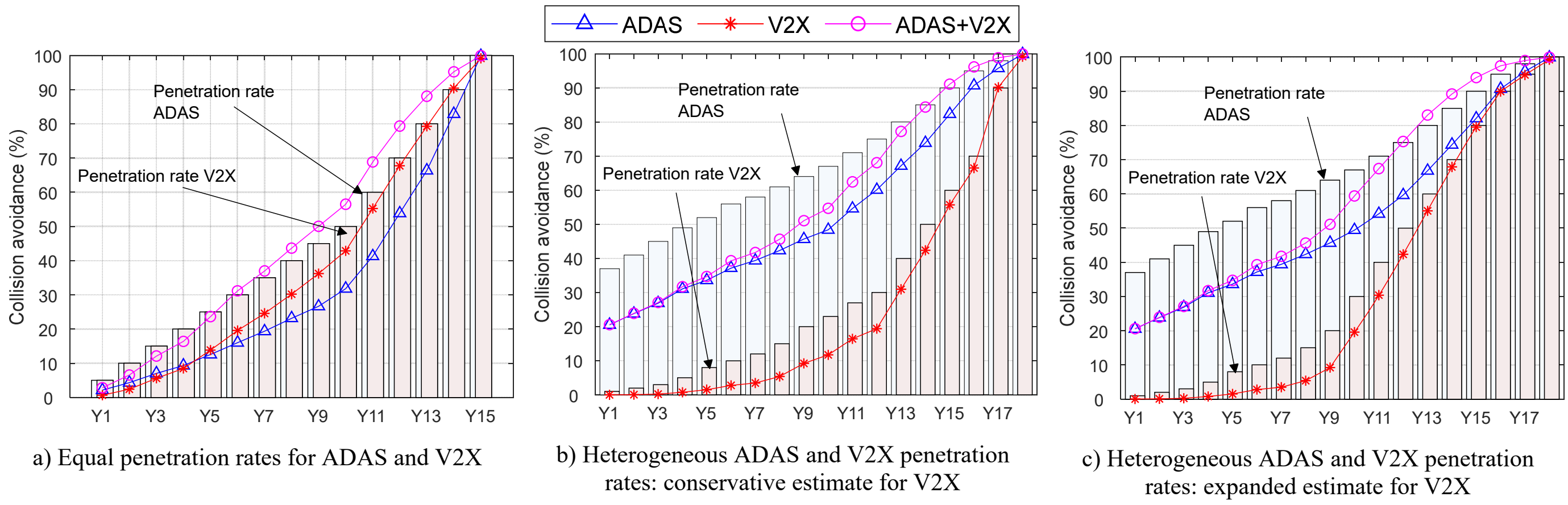

a) Equal penetration rates for ADAS and V2X

b) Heterogeneous ADAS and V2X penetration rates: conservative estimate for V2X

c) Heterogeneous ADAS and V2X penetration rates: expanded estimate for V2X

Fig. 2. Reduction in collisions (in percentage) as a function of the technologies' penetration rate over different years compared to the scenario where vehicles do not have ADAS or V2X.

introducing V2X still enhances vehicle safety and reduces collisions even at low penetration rates. Significant gains are obtained from combining ADAS and V2X once V2X penetration reaches about 20-30%. The comparison of Fig. 2.b and Fig. 2.c further show that accelerating V2X rollout considerably augments the safety benefits of the joint deployment of ADAS and V2X.

The benefits of jointly deploying ADAS and V2X is also visible when analyzing accident severity metrics for those collisions that could not be avoided. To this aim, we measure the relative speed between vehicles at the moment of collision ($\Delta v$). We classify collisions based on the relative speed into three severity levels: low ($\Delta v \leq 15$ m/s), medium ($15 < \Delta v \leq 30$ m/s), and high ($\Delta v > 30$ m/s). Fig. 3 presents the percentage of collisions in each severity category relative to the total number of collisions at Y6 and Y12 of the *expanded* deployment scenario. The penetration rates of ADAS and V2X are 56% and 10%, respectively for Y6, and 75% and 50% respectively for Y12. The results indicate that ADAS alone reduces accident severity with respect to the unequipped scenario where vehicles do not have ADAS or V2X, as evidenced by the lower proportion of high-severity collisions while increasing collisions with medium and low severity. Moreover, the combined deployment of ADAS and V2X not only decreases the total number of collisions compared with ADAS alone (Fig. 2), but also maintains or further reduces the percentage of high-severity crashes.

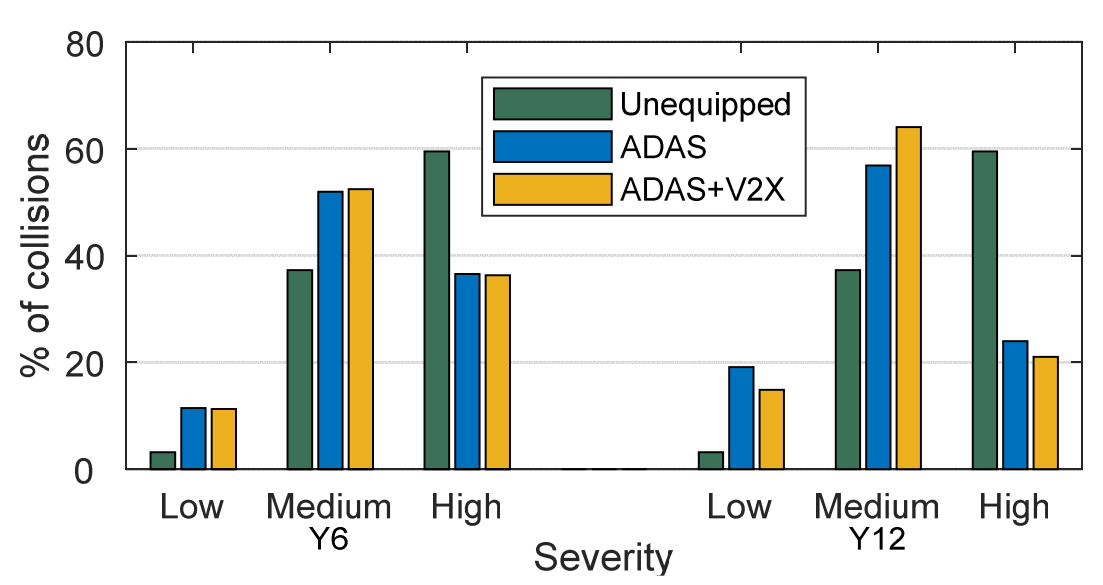

Fig. 3. Percentage of collisions with high, medium and low severity for Y6 and Y12 under the *expanded* deployment strategy.

Fig. 4 shows, for vehicles avoiding a collision after an emergency braking maneuver, the average stopping-distance margin to the collision, i.e. the distance between the ego vehicle after coming to a complete stop and the preceding vehicle. Results are depicted for the V2X *expanded* deployment strategy but similar trends were observed for the *conservative* deployment case. This figure only considers situations that do not result in a collision even in the *Unequipped* scenario. The results indicate that the joint deployment of ADAS and V2X provides a larger stopping-distance margin than deploying either technology alone across all evaluated penetration rates. The combined ADAS and V2X deployment enables earlier detection of the emergency situation and allows vehicles to brake sooner than vehicles only equipped with ADAS. The larger stopping-distance margin allows vehicles to stop under safer conditions, finally increasing vehicle safety.

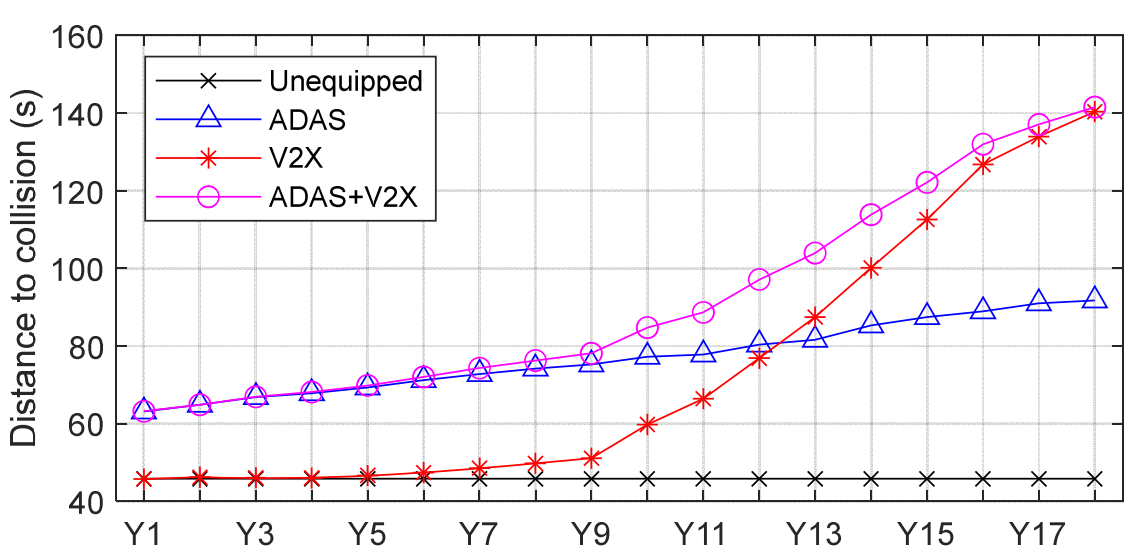

Fig. 4. Average stopping-distance margin to collision in the *expanded* deployment strategy. Each year has a different penetration rate of ADAS, V2X, and ADAS+V2X (Fig. 2.c).

## V. Conclusions

This paper analyzes the impact of ADAS and V2X penetration rates on safety in emergency-braking scenarios. The results show that both technologies individually increase vehicular safety and reduce the probability of collision, reduce the severity of collisions, and improve the safety conditions under which collisions are avoided. However, the study shows that deploying ADAS and V2X together combines the strengths of both, yields greater safety improvements, and requires lower penetration rates to achieve safety gains compared to only deploying ADAS or V2X. These trends are amplified with faster deployment of V2X under realistic scenarios. Future work will examine other use cases, such as overtaking maneuvers and intersections, where combined deployment is expected to deliver even greater benefits due to possible ADAS limitations under blockage conditions.